\documentstyle[12pt]{article}

\newcommand{\be}{\begin{equation}}
\newcommand{\ee}{\end{equation}}
\newcommand{\beqar}{\begin{eqnarray}}
\newcommand{\eeqar}{\end{eqnarray}}

\newcommand{\Tr}{{\rm Tr}}
\newcommand{\half}{{\frac{1}{2}}}

\expandafter\ifx\csname mathbbm\endcsname\relax

\else

\fi
\begin{document}
\begin{titlepage}
\begin{flushleft}
       \hfill                      CCNY-HEP-01-09\\
       \hfill                      RU-01-17-B\\
       \hfill                      November 2001\\
\end{flushleft}
\vspace*{3mm}
\begin{center}
{\LARGE Spectrum of Schr\"odinger field in a noncommutative magnetic monopole
\\}
\vspace*{12mm}
{\large D. Karabali$^a$, V.P. Nair$^b$ and 
A.P. Polychronakos$^{c,d}$\footnote{On leave from Theor.~Physics Dept., Uppsala University,
Sweden.
\hfil\break
E-mail: karabali@alpha.lehman.cuny.edu, vpn@sci.ccny.cuny.edu,
poly@teorfys.uu.se}
\\
\vspace*{5mm}
{\it a. Department of Physics and Astronomy \\Lehman College of the CUNY,
Bronx, NY 10468, USA}\\
\vspace*{4mm}
{\it b. Physics Department, City College of the CUNY \\
New York, NY 10031, USA}\\
\vspace*{4mm}
{\it c. Physics Department, The Rockefeller University \\
New York, NY 10021, USA}\\
\vspace*{4mm}
{\it d. Physics Department, University of Ioannina \\
45110 Ioannina, Greece}\\}

\vspace*{10mm}
\end{center}
%\maketitle

\begin{abstract}
The energy spectrum of a nonrelativistic particle on a noncommutative
sphere in the presence of a magnetic monopole field is calculated. 
The system is treated in the field theory language, in which the 
one-particle sector of a charged Schr\"odinger field coupled to
a noncommutative $U(1)$ gauge field is identified. It is shown that the
Hamiltonian is essentially the angular momentum squared of the particle, 
but with a nontrivial scaling factor appearing, in agreement with the 
first-quantized canonical treatment of the problem. Monopole quantization 
is recovered and identified as the quantization of a commutative 
Seiberg-Witten mapped monopole field.
\end{abstract}

\end{titlepage}

\section{Introduction}

Noncommutative spaces arise in string theory and in the matrix
model description of M-theory \cite{BFSS,CDS,SW}. Gauge and field
theories on such spaces have been extensively studied. For recent
reviews, see \cite{Har,DN,Sza}.

The quantum mechanics of particles on noncommutative spaces
has also been considered in some detail
\cite{Nai,NP,GLR,MP,DH,BNS,HK,IR}.  In particular, the problem of a
charged particle on the noncommutative plane \cite{NP} and the
noncommutative sphere \cite{Nai,NP} in an external constant magnetic
field has been analyzed using the canonical approach, in which
spatial noncommutativity relations complement the usual quantum
mechanical commutation relations.                    

In field theory, particles emerge as field quanta. It is, therefore,
of interest to recover the quantum mechanics of (non-interacting)
particles as excitations of fields on noncommutative spaces with
quadratic action. This would amount to obtaining and solving an equation
on a noncommutative space for a Schr\"odinger-like field coupled to
electromagnetic potentials. In \cite{NP} it was described how this
can be done for the noncommutative plane, and was pointed out that
the two approaches (canonical and field theory) are equivalent.

For the noncommutative sphere with a constant magnetic field,
however, only the canonical approach was carried out.
Although the canonical structure of the problem is
quite clear, certain subtleties arise when one considers the limit
in which the radius of the sphere goes to infinity, in relation
to the correspondence with the planar results. Specifically, the
Hamiltonian of the particle is not simply the square of the angular
momentum, as one may naively expect; a magnetic field-dependent
factor must be introduced in order to recover the correct planar
limit. Further, monopole quantization (which is a purely group
theoretic effect) does not imply the quantization of the strength
of the magnetic field itself $B$, but rather of an effective magnetic
field ${\tilde B} = B / (1-\theta B)$. Both results clearly call
for an explanation.

In this letter we give the field-theory treatment of the problem
of the particle on the noncommutative sphere, by coupling a charged
scalar field to a noncommutative monopole gauge field. The
Hamiltonian is simply defined as the covariant kinetic term of this
field. In the process, the scaling factor and the modified monopole
quantization naturally appear and the correct planar limit is obtained,
including even the zero-point energy of the ground state.

\section{Review of $U(1)$ gauge theory on the noncommutative sphere}

The noncommutative (NC) or `fuzzy' sphere has been introduced a while
ago \cite{Mad} and field theories on this space have been studied
rather extensively \cite{CW,Kli,Gra,IKTW}. We give here a self-contained
exposition of the basic notions of gauge theory on the NC sphere using
the operator (matrix) formulation.

We shall consider a noncommutative sphere defined in terms of the
three coordinates $x_i$, satisfying
\be
[ x_i , x_j ] = i \frac{\theta}{r} \epsilon_{ijk} x_k
\label{CSX}
\ee
$\theta$  and $r$ are real parameters which, without loss of generality,       
can be chosen positive. (A negative $\theta/r$ corresponds to a
parity transformation of the above space.) They will shortly
be identified with the noncommutativity parameter and the
radius of the sphere.

Clearly $R_i = (r/\theta) x_i $ satisfy the $SU(2)$ algebra. We shall
choose the $R_i$ to be in an irreducible representation of
spin $j$. $R_i$ and $x_i$ become matrices of dimension
$N = 2j+1$. Requiring $x_i^2$ to be the square radius of space, namely $r^2$,
fixes the scaling of $\theta$
\be
\theta = \frac{r^2}{\sqrt{j(j+1)}}
\label{ascal}
\ee
Near the ``north pole''
(that is, for states near the highest spin state of the
representation) we have $x_3 \simeq r$, so $[ x_1 , x_2 ]
\simeq i \theta$. We can, therefore, identify $\theta$ above
as the noncommutativity parameter of the space.
Slightly different definitions with the same scaling as (\ref{ascal})
for $j \to \infty$ are also acceptable. The definition
\be
\theta' = \frac{r^2}{j+\half} = \frac{4\pi r^2}{2\pi N}       
\label{thetaprime}
\ee
in particular, identifies $2\pi \theta'$ as the elementary area
on the sphere
corresponding to each of the $N$ states of the model, therefore
making the allusion to phase space quantum mechanics clearer.
We shall keep (\ref{ascal}) as our definition of $\theta$.

The two parameters $r$ and $j$, or, equivalently, $r$ and $\theta$,
define the model. Both the commutative sphere and the noncommutative
plane are obtained as the limit $j \to \infty$ but with different
scalings of $r$. Taking $r$ fixed, $\theta \to 0$, that is,
$j \to \infty$, recovers the commutative sphere, while taking
$r \to \infty$, $\theta$ fixed, that is, $j \to r^2/\theta$
recovers the noncommutative plane.

We shall consider field theory on the noncommutative sphere
with a commutative time.
Scalar fields on the NC sphere correspond to arbitrary $N \times N$
matrices whose elements are functions of time.
Such matrices can always be expanded in terms of monomials
of $x_i$ with any fixed ordering (such as the symmetric ordering).
So they can be thought of as functions of the noncommutative
coordinates of the sphere.
The fact that the Casimir $x^2 = r^2$ is fixed reduces the number of
independent functions in the same way as in the commutative
sphere. The fact that $x_i$ are of rank $N$, on the other hand,
and therefore $x_i^N$ can be reduced to lower-order polynomials
of the $x_i$ is a purely nonperturbative effect on the NC sphere
and amounts to a cutoff to the possible `Fourier modes' (that is,
spherical harmonics) of the allowed functions, reflecting the
finiteness of the Hilbert space of these functions.

The matrices $R_i = (r/\theta) x_i $ satisfy
\be
[ R_i , x_j ] = i \epsilon_{ijk} x_k
\label{angm}
\ee
Therefore, the adjoint action (commutator) of $R_i$ on any function
$\psi$ of $x_i$ generates spatial rotations of the $x_i$ and can be identified
with the angular momentum $L_i$:
\be
L_i (\psi) \equiv [R_i , \psi ]
\ee
Correspondingly, in analogy with the operator formulation of the
NC plane, the operators          
\be
\partial_i = -\frac{i}{\theta} x_i = -\frac{i}{r} R_i
\ee
can be identified with partial derivatives. Integration over the NC
sphere amounts to taking the trace with respect to the matrix space:
\be
\int \psi = \frac{4\pi r^2}{N} ~\Tr \psi
\label{intTr}
\ee
The prefactor in (\ref{intTr}) is inserted to achieve compatibility
with the proper commutative measure of integration; it ensures,
for instance, that the area of the sphere is recovered as
$\int 1 = 4\pi r^2$.

Gauge fields are introduced by deforming the above derivative
operators into covariant derivatives
\be
D_i = \partial_i + i A_i ~,~~~ D_0 = \partial_0 + i A_0
\ee
where $A_i$ and $A_0$ are time-dependent hermitian $N \times N$
matrices. The $A_i$ play the role of
the three spatial components of the gauge field in the ambient three
dimensional space of the sphere.
In the above, $\partial_0$ is an ordinary differential operator while
$\partial_i$ are $N \times N$ matrices. The $D_i$ are, thus, arbitrary
time-dependent anti-hermitian matrices.
Gauge transformations become conjugations of $D_i$ and $D_0$ by the
same arbitrary $U(N)$ matrix and correspond to a time-dependent
change of basis for the $N$ states on which the above matrices act.

The gauge field on a sphere has only 2 independent spatial
components, since the radial component is missing.
In the commutative case this is expressed by the relation
\be
x_i A_i = 0
\label{constrA}
\ee
which is also supplemented by the relation
\be
x_i L_i = 0
\label{constrL}
\ee
expressing the fact that $L_i$ are derivatives along the surface
of the sphere and do not act in the radial direction. To find the  
generalization of the above formulae in the NC case we note that
from the covariant $D_i$ we can construct new covariant coordinate
fields $X_i = i \theta D_i$. In the absence of gauge fields, $X_i$ are
the undeformed coordinates $x_i$. In general, they can be interpreted
as the NC coordinates of a membrane in the matrix regularization.
The absence of a third component of the gauge field
corresponds to the absence of radial excitations of the membrane;
the membrane should remain spherical, that is,
\be
X_i^2 = r^2
\ee
In terms of the covariant derivative matrices $D_i$, this becomes
\be
D_i^2 = -\frac{r^2}{\theta^2} = -\frac{j(j+1)}{r^2}
\label{restrD}
\ee
The above relation is obviously gauge invariant and replaces
({\ref{constrA}) in the NC case. Expressing $D_i = \partial_i
+ i A_i = -(i/\theta) x_i + i A_i$ we obtain
\be
x_i A_i + A_i x_i - \theta A_i^2 = 0
\ee
which becomes (\ref{constrA}) in the limit $\theta \to 0$, where $x_i$ and 
$A_i$ commute. (\ref{constrL}), on the other hand, takes the form
\be
x_i L_i (\psi) + L_i (\psi) x_i = x_i [ R_i , \psi ] + [R_i , \psi ] x_i = 0
\ee
which, upon putting $R_i = (r/\theta) x_i$ and using $x_i^2 = r^2$,
becomes an identity.

Since $[ \partial_i , \partial_j ] - \frac{1}{r} \epsilon_{ijk} \partial_k=0$,
the field strength $F_{ij}$ is defined as
\be
i F_{ij} = [ D_i , D_j ] - \frac{1}{r} \epsilon_{ijk} D_k
\ee
For $D_i = \partial_i$, the above definition gives $F_{ij} = 0$, as it should.
$F_{ij}$ is obviously gauge covariant under a gauge transformation,
\be
D_i \to U D_i U^{-1} \Longrightarrow F_{ij} \to U F_{ij} U^{-1}
\ee
In the planar limit $r \to \infty$, $iD_3 \to r/ \theta$, $iF_{ij}$ goes
to $[ D_i , D_j] + (i/\theta)$ which is the standard planar expression.
The electric components of the field strength $F_{0i}$ are, as usual,
\be
F_{0i} = -i [ D_0 , D_i ] = \partial_0 A_i - [ D_i, A_0 ]
\ee 
{}From $F_{ij}$ we can define a magnetic field
\be
i B_i = i \half \epsilon_{ijk} F_{jk} = \epsilon_{ijk} D_j D_k - \frac{1}{r} D_i
\label{Bfield}
\ee
A gauge covariant magnetic strength can now be defined
in analogy to the commutative one as
\be
B = \frac{X_i}{r}  B_i = i\frac{\theta}{r} D_i B_i = \frac{\theta}{r}
\epsilon_{ijk} D_i D_j D_k  + \frac{1}{\theta}~,
\ee
with a corresponding gauge invariant total flux over the sphere
\be
\Phi = \int B = \frac{4\pi r \theta}{N} \epsilon_{ijk} \Tr (D_i D_j D_k )
\, + \, \frac{4\pi r^2}{\theta}
\label{Phi}
\ee
Note, however, that the above total flux is neither topologically conserved
nor an integer multiple of $2\pi$. The `correction'
$4 \pi r^2/\theta = 4\pi \sqrt{j(j+1)}$ is
almost $2\pi N$ and an appropriate definition of $\theta$ (different from
either (\ref{ascal}) or (\ref{thetaprime}) above) would make it exactly
$2\pi N$.

\section{Monopole fields on the noncommutative sphere}

The gauge field of a magnetic monopole at the center of
the sphere would correspond to
\be
B_i = \frac{B}{r} X_i = \frac{i\theta B}{r} D_i
\label{Bmonopole}
\ee
with $B$ a constant number. This is the gauge-invariant counterpart
of the corresponding relation for a commutative abelian monopole field.
Using (\ref{Bfield}) the above relation becomes
\be
[ D_i , D_j ] = \frac{1-\theta B}{r} \epsilon_{ijk} D_k
\label{Dmonopole}
\ee
For $\theta B \neq 1$, this means that
$r D_i /(1-\theta B)$ satisfy the $SU(2)$ algebra.
If the $D_i$ are irreducible (that is, they cannot be brought to a
block-diagonal form with an appropriate gauge transformation),
since they are of dimension $N=2j+1$, they have to be proportional
to $\partial_i$ or $x_i$ (or any gauge transformation of the above).
Due to relation (\ref{restrD}) which fixes their overall scale, we conclude
that $D_i$ have to be equal to $\partial_i$ which implies $B=0$.
For $\theta B = 1$, on the other hand, the above relation implies
that all $D_i$ commute, and they are certainly completely reducible.

The only possibility, therefore, is to have reducible $D_i$. This means,
however, that as far as the action of covariant derivatives is concerned,
space has decomposed into non-communicating components, corresponding
to the blocks. In the membrane picture, the membrane represented by the $D_i$
consists of several concentric spherical components of the same radius
without any connection between them. In the degenerate case of one-dimensional
blocks (corresponding to $\theta B = 1$) the corresponding membranes are
single points at the surface of a sphere of radius $r$. As a result, the
dynamics
of a scalar wavefunction corresponding to reducible $D_i$'s decomposes
into non-communicating sectors, each corresponding to a different sphere.
For an irreducible sector of dimension $M = 2k+1$, the commutation relation
(\ref{Dmonopole}) and normalization condition (\ref{restrD}) imply
\be
(1-\theta B )^2 = \frac{j(j+1)}{k(k+1)} \equiv \gamma^2
\label{Bk}
\ee
where we defined the dimensionless factor $\gamma = 1-\theta B$.
Therefore, the monopole field $B$ in each of these sectors is determined by 
the dimensionality of the sector. There is no need to choose the same monopole
field $B$ in each of these sectors since they are not communicating, and we can
restrict ourselves in one of these sectors of dimension $M = 2k+1$
corresponding
to a desired $B$ field. This is the central observation, which will be further
justified when we analyze the Schr\"odinger equation in the next section.

In conclusion, for a monopole field we can write the $D_i$ in a block-diagonal
form: the upper component $D_i^u$, of dimensionality $M = 2k+1$, is of the form
\be
D_i^u = \frac{1-\theta B}{r} K_i = \frac{\gamma}{r} K_i
\label{Dred}
\ee
with $K_i$ irreducible $SU(2)$ matrices of spin $k$. In the above we used
relation (\ref{Bk}) for the monopole field $B$. $D_i^u$ obviously satisfies
(\ref{Dmonopole}) and (\ref{restrD}). The lower component $D_i^\ell$, of dimensi
onality
$M' = N-M$, corresponds to the decoupled part and can be any set of matrices
satisfying (\ref{Dmonopole}) and (\ref{restrD}). It could be, for instance,
a diagonal part with entries $-(i/\theta) z_{i,m}$, $m=1, \dots M'$.
This gauge field is of the projector type \cite{Pol,GN} and its field
strength corresponds to $M'$ `fuzzy Dirac strings' at positions $z_i$
on the sphere which can be thought of as bringing in the flux \cite{HaK}.
Alternatively, the lower component
could also be an irreducible space of the form (\ref{Dred}) for some
$k' = j-k-\half$, corresponding to a disconnected fuzzy sphere with
the `complementary' magnetic field.

It is clear from the above that there is a constraint on the
dimensionalities $M$ and $M'$ (summing to $N$) and a `monopole quantization'
corresponding to the fact that $2k$ is always an integer. This, however,
does not translate into the quantization of the conventional flux $\Phi$
as defined in (\ref{Phi}). To get a feeling of what this monopole quantization
corresponds to, we solve (\ref{Bk}) for $B$ and calculate the quantity
${\tilde B} = B/(1-\theta B)$:
\be
{\tilde B} = \frac{B}{1-\theta B} = \frac{1}{r^2} \left(
\pm \sqrt{k(k+1)} - \sqrt{j(j+1)} \right)
\label{Btil}
\ee
In either the commutative sphere limit or the noncommutative plane limit,
$j \to \infty$.  If we want the total flux $\tilde \Phi$ of the field
$\tilde B$ through
the surface of the sphere to remain finite, we must choose the positive sign
in (\ref{Btil}) and take the difference $k-j$ to remain finite. 
In that limit we obtain
\be
{\tilde \Phi} = 4\pi r^2 {\tilde B} = 2\pi \cdot 2(k-j)
\ee
Thus the flux $\tilde \Phi$ is quantized in integer multiples of $2\pi$, which
is the standard monopole quantization.

That $\tilde B$ rather than $B$ enters the quantization condition is not so
surprising. It is $\tilde B$, for instance, which determines the density of
states for a charged particle in each Landau level on the noncommutative
plane \cite{NP}.
A related analysis on the noncommutative two-torus also revealed the
special role of $\tilde B$ \cite{MP}.
Finally, the Seiberg-Witten transformation \cite{SW}, which
maps a noncommutative gauge field to a corresponding commutative one,
actually maps a constant noncommutative field $B$ to a commutative one
$\tilde B = B/(1-\theta B)$. Since the gauge transformation properties of
the field $\tilde B$ are the usual commutative abelian ones, we expect
monopole quantization to hold for $\tilde B$.

One last remark is in order. Just as we have taken $k<j$ in the above,
we could formally take $k>j$ to obtain the opposite monopole number.
This cannot, of course, be obtained through  
a reducible representation of the $D_i$'s in the original space, but it
can be obtained as a representation in a larger space. The point
remains, in each case, that a monopole charge in a NC sphere corresponds
to {\it two} NC spheres, one with spin $j$ representing the uncharged sphere 
and one with different spin $k$ representing the charged sphere, with
their difference fixed by the monopole number.

We conclude by mentioning that monopoles in the NC sphere and the
associated Dirac operators have been studied in \cite{GKP,BBVY}
using sections of bundles or partial isometries approach. Our analysis
in this section serves as a more conventional treatment in the standard
NC field theory language.

\section{Charged particle field on the NC sphere with a monopole}

We can now address the issue of the quantum states and energies of a
charged particle on the NC sphere with a magnetic monopole field. We
shall consider time-independent magnetic field with a monopole charge
$n = 2(k-j)$ constructed as above and take $A_0 =0$.

In noncommutative gauge theory, matter fields can only be in the fundamental
($F$), antifundamental ($\overline F$) or adjoint ($A$) representation.
The origin of this restriction is
very simple and has to do with the possible couplings of a field to the gauge
covariant operators $D_\mu$. A matter field $\psi$ is represented as a matrix in
noncommutative space. The only possible couplings to the operators $D_\mu$
which preserve the matrix structure are matrix multiplications. Considering,
then, kinetic terms that reduce to the proper derivative (commutator) form
$\partial_i (\phi) = [\partial_i , \phi ]$ in the zero gauge field limit,
we have the possibilities,
\beqar
F:~~D_i (\psi) &=& D_i \psi - \psi \partial_i ~,~~~~ \psi \to U \psi  \\
{\overline F}:~~D_i (\psi) &=& \partial_i \psi - \psi D_i
~,~~~~ \psi \to \psi U^{-1} \label{coupl} \\
A:~~D_i (\psi ) &=& D_i \psi - \psi D_i  \,,~~~\, \psi \to U \psi U^{-1} 
\label{adj}
\eeqar
In the $U(1)$ case, the adjoint matter field would decouple in the commutative
limit and does not correspond to a charged field. The fundamental and antifundam
ental
cases correspond to fields with charges $+1$ and $-1$ in natural units. Their
treatment and corresponding spectra are essentially identical. For our analysis
we shall choose the antifundamental case, to conform with the conventions 
of \cite{NP}
and to make the monopole field above (with monopole number $n = 2(k-j) <0$)
a positively charged one.

We consider a Lagrangian for $\psi$ of the Schr\"odinger type:
\be
{\cal L} = \frac{4\pi r^2}{N} \Tr \left( i \psi^\dagger {\dot \psi}
+ \half D_i (\psi )^\dagger D_i (\psi) \right)
\label{Sc}
\ee
As usual, the one-particle sector of the field theory defined by this
Lagrangian is described by a wavefunction $\psi$ which satisfies the
Schr\"odinger equation implied by (\ref{Sc}).
The corresponding energy eigenvalue problem is
\be
H (\psi) = -\half D_i^2 (\psi ) = E \psi \label{eigen}
\ee
where $D_i (\psi )$ is defined in (\ref{coupl}).

In the covariant derivative (\ref{coupl}) above, $\partial_i$ are
irreducible $N \times N$ matrices, but $D_i$ are reducible block-diagonal
matrices. As a result, $D_i (\psi )$
decouples into two non-mixing components. We decompose $\psi$ into two
rectangular components
\be
\psi = ( \phi | \phi' )
\ee
where $\phi$ is a $M \times N$ matrix, representing the first $M$ columns
of $\psi$, while $\phi'$ is a $M' \times N$ matrix representing the 
remaining $M'$ columns.
The covariant derivative of $\psi$ then takes the form
\be
i D_i (\psi ) = \left(\left. iD_i^u ( \phi ) ~\right| iD_i^\ell (\phi') \right)
= \frac{1}{r} \left( R_i \phi - \gamma \phi K_i ~\right|
\left.  R_i \phi' - \gamma' \phi' K_i' \right)
\label{Dd}\ee
So $D_i (\psi )$ breaks into two decoupling components, the one
of interest for our problem $\phi$ and the complementary one $\phi'$.
The kinetic energy term of $\psi$, $-\half D_i^2 (\psi )$, appearing in
the Schr\"odinger equation, similarly decomposes
into two decoupling components, one for $\phi$ and one for $\phi'$. The
particle could be either in the `upper' sphere, with wavefunction $\phi$
or in the `lower'
one with wavefunction $\phi'$. Although we could consider linear superpositions
of these states, there is nothing that could mix the two components and
therefore we may choose as a superselection rule $\phi' =0$.
This justifies the restriction
to the upper component of space alone, as argued in the previous section.
On this component we have
\be
i D_i (\phi ) = \frac{1}{r} R_i \phi - \frac{\gamma}{r} \phi K_i
\label{Dp}\ee
We note that the matrix $R_i$ acts on $\phi$ on the left, while the matrix
$-K_i$ acts on the right. As a result, when acting on $\phi$, $R_i$ and $K_j$
commute. Note, further, that $-K_i$ when acting on the right of $\phi$,
satisfy $SU(2)$ commutation relations.
We can therefore think of the $M \times N$ degrees of freedom in the matrix
$\phi$ as spanning the tensor product of two spaces $V_R$ and $V_K$ on
which $R_i $ and $K_i$ act, $V_R$ being the $N$ states of the spin-$j$
representation of $SU(2)$ carried by $R_i$ and $V_K$ being the $M$ states
of the spin-$k$ representation carried by $K_i$. So, the matrix elements
$\phi_{mn}$ (which parametrize the quantum state of the particle)
go over to a state $|\phi \rangle$
\be
|\phi \rangle = \sum_{n,m} \phi_{mn} |k,m \rangle \otimes |j,n \rangle
\ee
with $|j,n \rangle$, $n=-j, \dots j$, being the states in a spin-$j$ multiplet.
Altogether, the covariant derivative $D_i (\phi)$ can
be written as
\be
iD_i |\phi \rangle = \frac{1}{r} (R_i + \gamma K_i ) |\phi \rangle
\ee
where, now, $K_i$ is the operator acting on $|\phi >$ corresponding to $-K_i$
acting by right-multiplication on the matrix $\phi$.

We have therefore made contact with the setting of \cite{NP}. The $R_i$
and $K_i$ above are the corresponding operators defined there in the
canonical framework. The new element, however, is the appearance of
the scaling factor $\gamma$ in the covariant derivative above. We also note
that the covariant angular momentum operators on $| \phi \rangle$,
defined as the operators $J_i$ which satisfy $SU(2)$ commutation relations
and commute as in (\ref{angm}) with $x_i$, are
\be
J_i (\phi ) = R_i \phi - \phi K_i ~,~~~~ J_i |\phi \rangle = (R_i + K_i ) |\phi
\rangle
\label{J}\ee
as in \cite{NP}. This can also be identified from the invariance
of the Schr\"odinger action. Consider the transformations generated by
\be
\delta \phi = i\omega_j ( R_j \phi - \phi K_j )
\label {df}
\ee
with $\omega_j$ time-independent real numbers.
The covariant derivative (\ref{Dp}) on $\phi$ transforms as
\be
\delta D_i (\phi ) = i \omega_j \left\{ R_j D_i (\phi) - D_i (\phi) K_j
- \epsilon_{ijk} D_k (\phi) \right\}
\label{deltaD}\ee
It can be checked that the Schr\"odinger Lagrangian (\ref{Sc}) is invariant 
under the above transformation. The Noether charge derived from this
symmetry is
\be
Q = \omega_i Q_i = \omega_i \frac{4\pi r^2}{N} \Tr \left[
\phi^\dagger (R_i \phi - \phi K_i ) \right]
\ee
and corresponds to the expectation value of the operator $J_i$ as defined
in (\ref{J}).

We can finally calculate the spectrum of the Hamiltonian $H = -\half D_i^2$.
On states $|\phi \rangle$ the above Hamiltonian takes the form
\be
H = \frac{1}{2 r^2} ( R_i + \gamma K_i )^2 = \frac{1}{2 r^2} \left(
\gamma (R_i + K_i )^2 + (1-\gamma) R_i^2 - \gamma(1-\gamma) K_i^2 \right)
\ee
Using the expressions for the Casimirs $R_i^2 = j(j+1) = r^4 /\theta^2$
and $K_i^2 = k(k+1) = \gamma^{-2} R_i^2$ we finally obtain
\be
H = \frac{\gamma}{2 r^2} \left\{ J^2 - \left(\frac{B r^2}{\gamma}\right)^2
\right\}
\label{Hf}\ee
where we used (\ref{Bk}) to express
\be 
\frac{B r^2}{\gamma} =
\frac{r^2}{\theta} \left( \frac{1}{\gamma} -1 \right) =
\pm \sqrt{k(k+1)} - \sqrt{j(j+1)}
\label{C}\ee
This is exactly the Hamiltonian identified in \cite{NP} (up to a
zero-point shift),
with the nontrivial scaling factor $\gamma$ naturally appearing.

What is more, the above Hamiltonian
even has the correct zero-point energy as compared with the planar limit.
To see this, note that $J_i = R_i + K_i$ is simply the sum of two
commuting $SU(2)$ operators and its Casimir $J^2$ takes values $\ell (\ell+1)$
with $\ell = |j-k| , \dots j+k$. The total angular momentum has a lowest value
equal to half the monopole number, as expected, and a highest value $j+k$
reflecting the finiteness of the NC Hilbert space. Its ground state is
$J_0^2 = |j-k| ( |j-k| + 1)$.
In the limit of the NC plane, both $j$ and $k$ go to infinity;
by (\ref{ascal}), $j$ becomes $r^2/\theta -\half$ in this limit,
while, by (\ref{Bk}), $k$ becomes $r^2/(\gamma \theta ) -\half$.
Their difference $j-k$ then becomes (taking $\gamma$ positive)
\be
|j-k| \to \frac{r^2}{\theta} \left| 1 - \frac{1}{\gamma} \right| =
\frac{|B| r^2}{\gamma}  
\ee
Putting everything together, the ground state of $H$ in (\ref{Hf}) becomes
\be
E_0 = \frac{|B|}{2}
\ee
which is the correct zero-point energy of the lowest Landau level.
The rest of the spectrum also matches, as demonstrated in \cite{NP}.

\section{Adjoint matter field}

One can similarly analyze the spectrum for an adjoint matter field. The energy
eigenvalue problem is the same as in (\ref{eigen}), but with $D_i(\psi)$
defined as in (\ref{adj}). $\psi$ is decomposed into
\be
\psi= \left( \matrix{\phi_1&\phi_2\cr
\phi_3&\phi_4}\right)
\ee
where $\phi_1$ is a $M \times M$ matrix, $\phi_2$ is a $M \times M'$ matrix,
$\phi_3$ is a $M' \times M$ matrix and $\phi_4$ is a $M' \times M'$ matrix. The
covariant derivative is of the form
\be
D_i \psi = \left( \matrix{[D_i^u, \phi_1]&D_i^u\phi_2-\phi_2D_i^l\cr  
D_i^l \phi _3-\phi _3 D_i^u&[D_i^l, \phi_4]}\right)
\ee
The corresponding kinetic energy term decomposes into four decoupled components
and
we can write
\be
H = - {1 \over 2} \left( D_L + D_R \right) ^2
\ee
where for the four $\phi$'s in $\psi$ we have the assignments
\beqar
\phi_1 & : & ~~D_L  =  {i \gamma K \over r},~~D_R = {i \gamma K
\over r}\cr
 \phi_2 & : &~~D_L  =  {i \gamma K \over r},~~D_R = {i \gamma ' K' \over
r}\cr
 \phi_3 & : &~~D_L  = {i \gamma ' K' \over r},~~D_R = {i \gamma K \over
r}\cr
 \phi_4 & : &~~D_L  = {i \gamma ' K' \over r},~~D_R = {i \gamma ' K' \over
r}
\eeqar
$\phi_1$ can be expanded as $\phi_1 = \sum \phi_{1nm} |k,n \rangle \otimes |k, m
\rangle$ with the energy eigenvalues
\be
E_1 = {\gamma^2 \over {2r^2}} \ell(\ell+1)
\ee
$l= 0,1, \cdots ,2k$. As $j,k \rightarrow \infty$, this coincides with the
spectrum of a free uncharged particle on the sphere. In the commutative limit
the adjoint field has no charge and so does not couple to the monopole.
The above result is as
expected. Similar results hold for $\phi_4$ as $j,k' \rightarrow \infty$. For
$k'=j-k-1/2$ finite and $j,k \rightarrow \infty$, the $\phi_4$ field has the
spectrum
\be
E_4 = {\gamma ^{'2} \over {2 r^2}} \ell(\ell+1)
\ee
$l=0,1,\cdots,2k'$. $E_4 \sim j(j+1)$ at large $j$ and the number of states
remains finite as $j,k \rightarrow \infty$, with $k'$ fixed. (Presumably
these modes become nonpropagating as $j \rightarrow \infty$.)
The ``monopole field" seen by the upper
components is proportional to $j-k$, the field seen by the lower components is
proportional to $j-k'$. Therefore if either is fixed and we take $j \rightarrow
\infty$, the field seen by the other becomes very large and this leads to the
above behavior of energy eigenvalues.

For the $\phi_2$ field we have
\beqar
E_2 & = &{{(\gamma K + \gamma ' K')^2} \over {2 r^2}} \cr 
& = & {{j (j+1)} \over {2r^2}} \left[ 2 - \sqrt{{{k(k+1)} \over {k' (k'+1)}}} -
\sqrt{{{k'(k'+1)} \over {k (k+1)}}} + {{(K + K')^2} \over
{\sqrt{k(k+1)k' (k'+1)}}} \right]
\eeqar
For finite monopole number, the lowest value of $E_2$ diverges as $j \rightarrow
\infty$, but the limit $j,k,k' \rightarrow \infty$ with $(k-k')$ fixed can give
finite energies. A similar statement holds for the $\phi _3$ field as well.

The adjoint fields behave in some respects like
combinations of the antifundamental representations suitably on the upper
and lower spheres. To see this, consider a field
$\psi$ which transforms as
$\phi
\otimes
\phi '$, where $\phi,~\phi'$ are in the antifundamental representations
as in section 4, $\phi$ on the upper sphere and $\phi'$ on the lower
sphere. $\psi$ may be expanded as
\be
\psi \sim \phi \otimes \phi ' = \sum \psi_{mnm'n'} |k, m \rangle \otimes
|j,n \rangle \otimes |k', m' \rangle \otimes |j, n' \rangle
\ee
The action of the
covariant derivative is, by using formula (\ref{Dd}),
\be
iD_i(\psi ) = (R_i + \gamma K_i + \gamma ' K'_i) \psi
\ee
$R_i$ acts on $|j,n\rangle$ and $|j, n' \rangle$, both being spin-$j$
representations. The singlet combination in $\phi \otimes \phi '$ with
$R_i =0$ behaves like $\phi_2$. One can think of
the other
components of the adjoint in a similar way.

\section{Two subtleties}

In the analysis of the previous sections we glossed over
two subtle points, which we wish to clarify here. They have to do
with the $\pm$ sign in (\ref{Btil},\ref{C}) and with the critical
magnetic field $\theta B = 1$.  

Equation (\ref{Bk}) for the magnetic field has two solutions, corresponding
to $\gamma >0$ and $\gamma <0$. This is the origin of the double signs
appearing in (\ref{Btil},\ref{C}). Thus, the same values of $j,k$
represent two values of the magnetic field, one subcritical $B_- <1/\theta$
and one overcritical $B_+ >1/\theta$,
corresponding to $\gamma = \pm |\gamma|$ and related by
\be
B_- + B_+ = \frac{2}{\theta}
\ee
The matrices $R_i$, $K_i$ describing the model are the same and the only
difference is the sign of the factor $\gamma$ appearing in the expressions
for $D_i (\phi)$ and $H$. To see how the two cases are imbedded in the
Hilbert space, notice that the Hamiltonian has a prefactor $\gamma$.
Thus, if $\gamma >0$ (corresponding to $B_- <1/\theta$) the ground
state of $H$ corresponds to the lowest eigenvalue of $J^2$ (as assumed
in the last paragraph of section 5). If, however, $\gamma <0$
(corresponding to $B_+ >1/\theta$) the ground state corresponds to the
highest eigenvalue of $J^2$. The analysis of the previous section
still goes through, with $j+k$, now, defining the ground state, and the
energy levels in terms of $B$ are as before. We conclude that
the energy spectrum for the two values of the magnetic field $B_\pm$ is
essentially the same but reversed, with the states at each end of the
spectrum mapping to two different magnetic fiels in the planar limit, 
as explained in \cite{NP}. Note, also, that for subcritical fields
the angular momentum for excited states increases, as expected, while
for overcritical fields the angular momentum {\it decreases} with energy.
(A similar effect for the planar case was pointed out in \cite{BNS}.)

There is a potential confusion with the critical field $B = 1/\theta$
corresponding to $\gamma =0$.  From (\ref{Bk}) it appears that $\gamma =0$
corresponds to the limit $k \to \infty$. On the other hand, from equation
(\ref{Dmonopole}) it appears that $\gamma =0$ corresponds to the completely
reducible case in which we should choose an irreducible $D_i = -(i/\theta) z_i$
with $z_i$ a real vector squaring to $r^2$. We now clarify the relationship
between these two cases.

There is a related issue in the planar case. In the first-quantized
picture, the algebra of observables for $\theta B =1$ becomes reducible.
An irreducible Hilbert space would put $x_i + \theta \epsilon_{ij} p_j$
($i,j=1,2$) equal to some numbers $z_i$, since they become Casimirs at the
critical value of $B$, and would correspond to a single Heisenberg copy
of the algebra of the $x_i$'s alone. The limit of the Hilbert space as
$B \to 1/\theta$, on the other hand, contains an infinite number of copies
of the irreducible space, embedded into the direct product of two Heisenberg
representations. An irreducible component in this space corresponds to
picking one state from each Landau level, centered around the
spatial point $z_i$ corresponding to the value of the Casimir.

On the sphere, the algebra of
observables is always irreducible and, a priori, there appears to be no mechanism
for recovering the irreducible planar Hilbert space for $B = 1/\theta$.
In fact, the reducibility of (\ref{Dmonopole}) achieves just that. The
representation obtained as the limit $k \to \infty$ maps to the limit
$B \to 1/\theta$ in the planar case, while the representation
$D_i = -(i/\theta) z_i$ maps to the irreducible planar Hilbert space.
To see this, note that the wavefunction $\phi$ for this choice of $D_i$
becomes a single copy of the spin-$j$ multiplet, on which $D_i$ acts as
\be
i D_i (\phi ) = \left( \frac{1}{r} R_i \phi - \frac{z_i}{\theta}\right) \phi
\label{Drred}\ee
The Hamiltonian becomes
\be
H = \half \left( \frac{1}{r} R_i - \frac{z_i}{\theta}\right)^2
= \frac{r^2}{\theta^2} - \frac{z_i R_i}{\theta r}
\ee
where we used $R_i^2 = r^4/\theta^2$ and $z_i^2 = r^2$. The matrix
${\hat R} = (z_i /r) R_i$ is clearly a spin-$j$ $SU(2)$ generator and
has eigenvalues ${\hat R} = j - n$, $n=0,1,\dots 2j$. So the energy
levels are
\be
E_n = \frac{n}{\theta} + \frac{r^2}{\theta^2} - \frac{j}{\theta}  
\ee
each with degeneracy one.

In the limit $r \to \infty$, $j$ becomes
$r^2/\theta - \half$. So the energy levels become
\be
E_n = \frac{1}{\theta} \left( n + \half \right)
\ee
These are exactly the Landau energy levels for $B = 1/\theta$,
with one state per level, as in the irreducible planar Hilbert space.
For these states the expectation value of $R_i$ is $\langle R_i \rangle = (z_i /r)
\langle {\hat R} \rangle$ and we get
\be
\langle x_i \rangle = \frac{\theta}{r} \langle R_i \rangle =
\frac{\theta z_i}{r^2} (j-n) ~\to ~ z_i - \frac{\theta z_i}{r^2}
\left( n+\half \right)
\ee
So in the limit $r \to \infty$ we get $\langle x_i \rangle = z_i$, as in the planar
case.

The reducibility of the system at $\theta B =1$ can be lifted if we
include the generators of spatial symmetries in the algebra of
observables. In the planar case, the Casimirs $x_i +
\theta \epsilon_{ij} p_j$ are not invariant under translations and
rotations. That is, the translation and rotation generators cannot be
expressed any more in terms of $x_i$ and $p_i$, and independent
generators for these translations have to be appended, rendering the
algebra irreducible and calling for the full Hilbert space of two
copies of the Heisenberg representation, as obtained in the
$B \to 1/\theta$ limit. This is the situation closest resembling 
the sphere. The full algebra of $x_i$ and $J_i$ is irreducible.
The reduced representation corresponding to (\ref{Drred}) breaks
rotational invariance (which would rotate the Casimirs $z_i$).
(Note that (\ref{deltaD}) does not hold if $\delta \phi$ is as in (\ref{df}) but
$D_i
\phi$ is as in (\ref{Drred}).) Introducing appropriate $J_i$ which also rotate the
$z_i$ calls for inclusion of representations with all possible $z_i$, reproducing
the infinite-dimensional Hilbert space as obtained in the $k \to \infty$
limit. 

In summary, the situation for $B = 1/\theta$ is similar between the
plane and the sphere except for the fact that on the sphere we already have the
$J_i$'s which can make changes (rotations) in $z_i$. On the plane we have to
enlarge the algebra of observables to incorporate shifts of $z_i$.

Let us also comment on the limit $k \to 0$, which
corresponds to $B \to \pm \infty$. In that limit,
the field $\phi$ becomes again a single copy of the spin-$j$ multiplet,
as in the reduced case $B = 1/\theta$ above. This could be surprising, since
the two cases represent vastly different values of $B$. What happens
in this case, in fact, is the well-known reduction of the Hilbert space
in a strong magnetic field to the lowest Landau level. The dimensionality
of the Hilbert space is $(2j+1)(2k+1)$. As $|B|$ becomes large, $k$
decreases, so the dimensionality of the space decreases. The density
of states per Landau level, on the other hand, increases linearly
with $|{\tilde B}|$ and becomes $\theta^{-1}$ for large $|B|$.
So the Hilbert space can accomodate fewer
and fewer Landau levels and, at the limit $k \to 0$, only one level
remains. The energy spectrum is quite different in the two cases.
For $k \to 0$ all $2j+1$ states become degenerate with a divergent
energy, while for the reduced $\theta B =1$ case they reproduce an
equidistant spectrum with spacing $\theta^{-1}$.

In summary, the NC sphere-cum-monopole presents quite a fascinating
`package': it captures all the features and special limits of planar
noncommutative theory, for two values of the magnetic field at a time,
all encoded in a finite Hilbert space.

\section{Conclusions}

We have analyzed the problem of a charged particle on the NC sphere from
the NC gauge field point of view. We identified the monopole sectors and
obtained the spectrum of a charged Schr\"odinger field both for the 
fundamental and adjoint couplings. Our results are in complete agreement with
the canonical analysis of the problem, and reproduce the nontrivial scaling
factor of the Hamiltonian, as well as the zero-point shift in the energy
required to obtain the
correct correspondence with the planar limit.  We also obtained the magnetic
monopole quantization and identified it as the  quantization of an equivalent
commutative monopole charge obtained through the Seiberg-Witten transformation.

Some other issues of the dynamics of a charged particle on the NC sphere 
in the presence of a monopole, such as inclusion of spin, are under
investigation.

\vskip .3in

\noindent
{\bf Acknowledgements}

We would like to thank B.~Morariu for discussions.
This work was supported in part by the National Science Foundation grants
PHY-9970724 and PHY-0070883 and PSC-CUNY-32 awards.

\end{document}